\documentclass{article}
\usepackage{epsfig}

\newcommand{\ceiling}[1]{ {\left\lceil #1 \right\rceil} }

\newcommand{\floor}[1]{ {\left\lfloor #1 \right\rfloor} }

\newcommand{\expect}[1]{ {\left\langle #1 \right\rangle} }

\newcommand{\tbl}[1]{Table~\ref{#1}}

\newcommand{\fig}[1]{Fig.~\ref{#1}}

\newcommand{\Pmin}{ {P_{\rm min}} }

\newcommand{\rhoInit}{ {\rho_{\rm init}} }
\newcommand{\rhoRate}{ {\rho_{\rm rate}} }
\newcommand{\Tmin}{ {T_{\rm min}} }

\newcommand{\Stotal}{ {S_{\rm total}} }
\newcommand{\tourlength}{ L }

\newcommand{\nBits}{ n }
\newcommand{\nSAT}{ \nBits }
\newcommand{\nTSP}{ N }
\newcommand{\Cost}{ \expect{C} }

\title{\bf Quantum Optimization}
\author{\bf Tad Hogg\thanks{to whom correspondence should be sent}\\
\small Xerox Palo Alto Research Center\\
\small 3333 Coyote Hill Road\\
\small Palo Alto, CA 94304\\
\small hogg@parc.xerox.com
\and 
\bf Dmitriy Portnov\\
\small Computer Science and Engineering Dept.\\
\small Univ. of Washington\\
\small Seattle, WA 98105\\
\small dportnov@cs.washington.edu 
}

\begin{document}
\maketitle

\begin{abstract}
We present a quantum algorithm for combinatorial optimization using
the cost structure of the search states. Its behavior is illustrated
for overconstrained satisfiability and asymmetric traveling salesman
problems. Simulations with randomly generated problem instances show
each step of the algorithm shifts amplitude preferentially towards
lower cost states, thereby concentrating amplitudes into low-cost
states, on average. These results are compared with conventional
heuristics for these problems.
\end{abstract}

%----------------------------------------------------------
\section{Introduction}

Quantum computers~\cite{deutsch85,divincenzo95} operate on
superpositions of all classical search states, allowing them to
evaluate properties of all states in about the same time a classical
machine requires for a single evaluation.  This property is known as
quantum parallelism.  Superpositions are described by a state vector,
consisting of complex numbers, called amplitudes, associated with the
classical states.

Most quantum search algorithms focus on decision problems, which have
an efficiently computable test of whether a given state is a solution.
Without using any information about the problems beyond this test,
quantum computers give a quadratic improvement in search speed by
using amplitude amplification~\cite{grover96,boyer96}.  Using more
information gives further improvement in some
cases~\cite{grover97b,hogg98,hogg00,spector99}, but it remains to be
seen how much improvement is possible for large, difficult search
problems.

Some combinatorial searches have so many desired properties for a
solution that none of the search states satisfy all of them, i.e.,
there is no solution. In such cases, one often instead asks for a
state with as many desirable properties as
possible~\cite{freuder92}. More generally, each state has an
associated cost and the goal is to find a minimum-cost state. Such
optimization searches can be treated as a series of decision problems
with different assumed values for the minimum cost. However many
classical heuristics for optimization problems find low-cost states
directly, although these are not guaranteed to be the actual
minimum. This raises the question of whether quantum algorithms can
show similar behavior since amplitude amplification does not directly
apply to optimization problems where the minimum cost is not known a
priori.

As a direct approach to optimizaton problems, this paper examines
algorithms mixing amplitudes among different states so as to gradually
shift the bulk of the amplitude toward states with relatively low
costs, a technique previously applied to a decision
problem~\cite{hogg00}. Like many classical methods, the resulting
quantum algorithms are heuristic (i.e., not guaranteed to find the
minimum-cost state) and incomplete (i.e., even if such a state is
found, the algorithm provides no definite indication that it is indeed
a minimum). In common with most studies of heuristic methods, we
evaluate their typical behavior on classes of problems rather than
determining worst-case bounds (which are often far more pessimistic
than typical behaviors). Specifically, the next section presents the
quantum algorithm in the context of a general optimization problem,
and contrasts it with amplitude amplification. The following two
sections then examine instances of the algorithm suitable for
overconstrained satisfiability problems (SAT) and asymmetric
traveling salesman problems (ATSP).

%----------------------------------------------------------
\section{Optimization Algorithm}

The quantum optimization algorithm presented here operates on
superpositions of all search states, and attempts to find a state with
relatively low cost. The cost associated with each search state is
used to adjust the phase of the state's amplitude, and a mixing
operation combines amplitudes from different states.

More specifically, the overall algorithm consists of a number of
independent trials, each of which returns a single state after the
final measurement. The number of trials can be fixed in advance if
some (hopefully low-cost) state is required within a preset time
bound, or can continue until some other criterion is satisfied, e.g.,
a sufficiently low cost state is found or a long series of trials
gives no further improvement. In this respect, this algorithm is
similar to incomplete classical heuristics which tend to give low-cost
states but do not guarantee to find the absolute minimum. Moreover,
even when the minimum is found, the algorithm offers no guarantee that
this is indeed the minimum cost so that further trials will not give a
lower cost.

\subsection{A Single Trial}

A single trial of the algorithm on a quantum computer with $\nBits$
bits to represent the search state consists of the following
efficiently implementable~\cite{boyer96,hogg98b} steps:
\begin{enumerate}
\item
   initialize the amplitude equally among the states, giving $\psi_s^{(0)} =
   2 ^ {-\nBits/2}$ for each of the
   $2^\nBits$ states $s$.

\item
   for steps 1 through $j$, adjust amplitude phases based on the costs associated with the states and then mix them. These operations correspond to matrix multiplication of the state vector, with the final state vector given by:
\begin{equation}\label{map}
\psi^{(j)} = U^{(j)} P^{(j)} \ldots U^{(1)} P^{(1)} \psi^{(0)},
\end{equation}
where, for step $h$, $U^{(h)}$ is the mixing matrix and $P^{(h)}$ is the
phase matrix, as described below.

\item
   measure the final superposition, giving state $s$ with
probability $p(s) = | \psi^{(j)}_s |^2$. Thus the probability to obtain a minimum cost state with a single trial is $\Pmin = \sum_s p(s)$ where the sum is over those $s$ with the minimum cost.
\end{enumerate}

The mixing matrix is $U^{(h)} = W T^{(h)} W$, where, for states $r$
and $s$, $W_{rs} = 2^{-\nBits/2}(-1)^{|r \wedge s|}$ is the Walsh
transform and $|r \wedge s|$ is the number of 1-bits the states have
in common.  The matrix $T^{(h)}$ is diagonal with elements depending
on $|s|$, the number of 1-bits state $s$ contains: $T^{(h)}_{ss} =
t^{(h)}_{|s|}$ with
\begin{equation}\label{tmatrix}
t^{(h)}_b = e^{i\pi \tau_h b}
\end{equation}
where $\tau_h$ is a constant depending on the class of problems and
the number of steps, but not the particular problem instance being
solved. From these definitions, the elements $U^{(h)}_{r s}$ depend
only on the Hamming distance between the states, $d(r,s)$, i.e., the
number of bits with different values in the two states. That is, we
can write $U^{(h)}_{r s} = u^{(h)}_{d(r,s)}$, with $u^{(h)}_d = ( -i
\tan(\pi \tau_h/2) )^d$, up to an overall phase and normalization
constant~\cite{hogg00}.

The phase adjustment matrix, $P^{(h)}$, is a unitary diagonal matrix
depending on the problem instance we're solving, with values
determined by the cost associated with each state: $P^{(h)}_{rr} =
p^{(h)}_{c(r)}$ and
\begin{equation}\label{phasematrix}
p^{(h)}_c = e^{ i \pi \rho_h c }
\end{equation}
where $\rho_h$ is a constant and $c(r)$ is the cost associated with
search state $r$.

This algorithm has the same overall structure as amplitude
amplification~\cite{grover96}. In fact, it reduces to amplitude
amplification if we define the ``cost'' of a search state to be 0 for
a solution and 1 otherwise and make the choices $t^{(h)}_0 = -1$,
$t^{(h)}_b = 1$ for $b>0$, $p^{(h)}_0 = -1$ and $p^{(h)}_1 = 1$ for
all steps $h$. Note that for optimization problems where the minimum
cost is not known a priori, none of the states will be solutions and
amplitude amplification gives no enhancement in the minimum-cost
states.  On the other hand, the multiple trials of this optimization
algorithm could be combined with amplitude amplification to achieve a
further quadratic improvement if the minimum cost were known or
through a series of repetitions using different assumed values for the
minimum~\cite{brassard98}.

\subsection{Applying the Algorithm}

Completing the specification of the algorithm requires the number of
steps $j$ and values for the phase parameters $\tau_h$ and $\rho_h$
for $h=1,\ldots,j$. We consider two approaches for identifying
parameters giving good performance. The first uses a sample from the
class of problems to be solved, and numerically adjusts the parameters
to give the largest probability of finding a minimum cost state when
averaged over the sample. This approach, commonly used to tune
classical heuristics, allows precisely tuning the parameters but is
limited to small problem sizes whose behavior can be simulated using
classical machines. Applying this approach to larger problems will
require the development of quantum hardware.

The second approach evaluates the asymptotic average behavior of the
algorithm, as a function of the phase parameters, and selects values
giving good average performance for large problems. When the number of
steps $j$ is held fixed as $\nBits$ increases, this can be done
exactly~\cite{hogg98e}. However, good performance requires the number
of steps to increase with the size of the problem, which complicates
this exact analysis. Instead, we can use an approximate evaluation of
the asymptotic behavior~\cite{hogg00}. In this approximation, the
amplitudes at each step are assumed to depend only on the costs
associated with the states. Let $\phi^{(h)}_c$ be the average
amplitude of states with cost $c$ after step $h$. With the above
definitions of the mixing and phase matrices, the change in average
amplitudes from one step to the next is approximately
\begin{equation}
\phi^{(h)}_{c'} = \sum_{d c} u^{(h)}_d p^{(h)}_c \phi^{(h-1)}_c \nu(c',d,c)
\end{equation}
where $\nu(c',d,c)$ is the average number of states with cost $c$ at
distance $d$ from a state with cost $c'$. This quantity can be
expressed as ${\nBits \choose d} P(c|d,c')$ where $P(c|d,c')$ is the
conditional probability a state has cost $c$ when at distance $d$ from
a state with cost $c'$. When a class of problems has a simple
expression for the asymptotic form of this conditional probability,
this approximate equation gives the behavior of the average
amplitudes. It can then be used to select phase parameters and the
number of steps to give a large enhancement in amplitudes for low-cost
states.

An optimization heuristc can be evaluated in a number of ways. For
example, $\Cost=j/\Pmin$ is the expected number of steps (including
repetitions due to multiple trials) required to produce a minimum-cost
state. Alternatively, one could ask how close the algorithm gets to
the optimum as a function of the number of trials. This latter measure
allows trade-offs between methods that give reasonably good results
very quickly, but then give little subsequent improvement, and those
that improve only slowly but eventually give lower cost
states. Finally, one could characterize a single trial by its
likelihood of returning the minimum cost, $\Pmin$, or the expected
cost of returned states, $\sum_c c \sum_{s | c(s)=c} p(s)$. In our
case we focus on $\Cost$ as a performance measure. However, since the
algorithms concentrate amplitude toward low-cost states, comparisons
based on the other measures give the same general conclusions.

%----------------------------------------------------------
\section{Satisfiability}

Satisfiability is a combinatorial search problem consisting of a
propositional formula in $\nSAT$ Boolean variables and the requirement
to find an assignment (true or false) to each variable so that the
formula is true.  For $k$-satisfiability ($k$-SAT), the formula is a
conjunction of $m$ clauses each of which is a logical OR of $k$
(possibly negated) variables.  In this form, every clause must be true
in order that the full formula is true.  A state (i.e., an assigned
value to each variable) is said to conflict with any clause it doesn't
satisfy. For $k\ge 3$, $k$-SAT is NP-complete~\cite{garey79}.  An
example 2-SAT problem with 3 variables and 2 clauses is ($v_1$ OR (NOT
$v_2$)) AND ($v_2$ OR $v_3$), which has 4 solutions, e.g., $v_1={\rm
false}$, $v_2={\rm false}$ and $v_3={\rm true}$.

$k$-SAT problems with many clauses typically have no solutions.  Such
cases give an optimization problem~\cite{freuder92}, namely to find
assignments with the minimum number of conflicts, i.e., the fewest
unsatisfied clauses.  To examine typical behavior of the algorithm, we
use the well-studied class of random $k$-SAT, in which the $m$ clauses
are selected uniformly at random. Specifically, for each clause, a set
of $k$ variables is selected randomly from among the $\nSAT
\choose k$ possibilities. Then each of the selected variables is
negated with probability $1/2$ to produce the clause. Thus each of the
$m$ clauses is selected, with replacement, uniformly from among the
${\nSAT \choose k} 2^k$ possible clauses. The difficulty of solving such
randomly generated problems varies greatly from one instance to the
next. This class has a high concentration of hard instances when $\mu
\equiv m/\nSAT$ is near a phase transition in search
difficulty~\cite{cheeseman91,kirkpatrick94,hogg96d}. For random 3-SAT
this transition is near $\mu=4.25$. For our study of optimization, we
generate these random problems but keep in the sample only those with
no solutions, as evaluated with a classical exhaustive search. The
minimum cost can vary among the instances in a sample.

\subsection{Algorithm}

Since SAT involves Boolean variables, the states in a problem with
$\nSAT$ variables can be directly represented with $\nSAT$ bits in a
quantum computer. With such a representation, the Hamming distance
between two bit-sequences corresponds to the number of variables
assigned different values in the two corresponding search states. This
correspondence allows a simple combinatoric expression for the
conditional probability $P(c|d,c')$ that an assignment has $c$
conflicts when at distance $d$ from another assignment with $c'$
conflicts. This expression can in turn be used to select algorithm
parameters that lead to significant shift of amplitude toward states
with few conflicts, on average~\cite{hogg00}.

Specifically, this approximation suggests using $j=\nSAT$ and a linear
variation in the phase parameters, i.e., $\rho_h = \frac{1}{j} (R_0 +
R_1 (1 - \frac{h-1}{j}))$ and $\tau_h = \frac{1}{j} (T_0 + T_1 (1 -
\frac{h-1}{j}))$. Thus, for instance, the first step uses
$\rho_1=(R_0+R_1)/j$. As $j$ increases, the $\rho$ and $\tau$ values
become small so the corresponding $P$ and $U$ matrices become close to
identity matrices, i.e., each step only introduces small changes in
the amplitudes.

The appropriate values of the four parameters, $R_0$, $R_1$, $T_0$ and
$T_1$, depend on the class of SAT problems, in particular the values
of $k$ and $\mu$ for random $k$-SAT. These values can be evaluated
numerically in the context of decision problems, i.e., attempting to
maximize the probability in solution states, assuming any
exist~\cite{hogg00}. This process gives parameter values suitable for
problems with solutions. In this paper, we apply the same values for
optimization problems.

\subsection{Behavior}

An example of the shift in amplitude is illustrated in \fig{SAT
example}.  The distribution for step 0 simply reflects the number of
states with each number of conflicts in the problem. Thus in this
example most assignments have about 10 conflicts.  Although this
problem has no solutions, it shows the same shift in amplitude toward
low-cost states as seen for soluble problems~\cite{hogg00}. In
particular, after the last step, the measurement is likely to produce
a minimum-conflict state. Furthermore, even if such a state is not
produced, the result is still very likely to have a relatively low
number of conflicts. Significantly, the algorithm operates even
without prior knowledge on the minimum number of conflicts. In fact,
the approximate theory used to numerically select the phase parameters
is based on maximizing the solution probability among soluble random
3-SAT with $\mu=4$ in this case. The figure shows such parameters also
work well for insoluble problems. Thus the correlation between
number of conflicts and Hamming distance used by the theory is roughly
the same for soluble and insoluble instances for most of the states.
Nevertheless, an open question is whether somewhat different phase
parameters may give better performance for optimization problems.

\begin{figure}[t]
\begin{center}
\epsfig{file=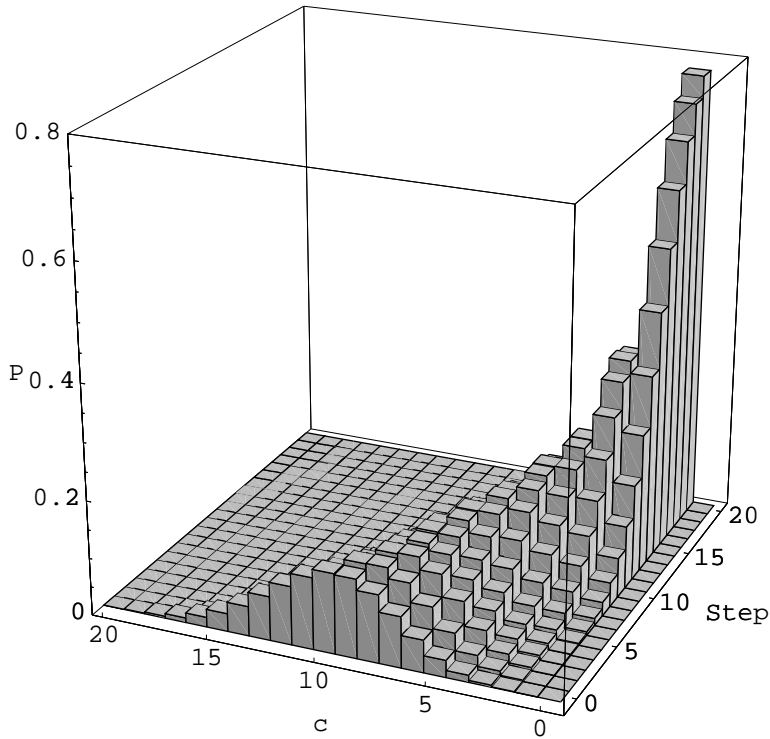,width=4in}
\end{center}
\caption{\label{SAT example}Probability to find an assignment with each number of conflicts vs.~number of steps for a 3-SAT problem with 20 variables, 80 clauses and no solution (so the probability of finding a state with 0 conflicts is always zero). In this case, the minimum number of conflicts is one. The algorithm used $T_0=0.539298$, $T_1=3.5105$, $R_0=4$ and $R_1=-3.4$ with $j=20$ steps.}
\end{figure}

Classical heuristics also often manage to find low-conflict states.
One such heuristic is GSAT~\cite{selman92}.  The GSAT algorithm starts
from a random assignment and, for each step, examines the number of
conflicts in the assignment's neighbors (i.e., assignments obtained by
changing the value for a single variable) and moves to a neighbor with
the fewest conflicts. If a solution isn't found after a prespecified
number of steps ($2 \nSAT$ for the comparison reported here), e.g.,
because the current assignment is a local minimum, the search is tried
again from a new random assignment.  As with the quantum algorithm,
GSAT is incomplete, i.e., does not guarantee its result is indeed a
minimum-conflict assignment. 

We use multiple trials to estimate the probability GSAT 
returns a minimum conflict state. Specifically, the
expected cost estimate is $\Cost = \Stotal / \Tmin$ where $\Stotal$ is
the total number of steps in all 1000 trials we used for each problem
and $\Tmin$ is the number of trials for which a minimum-conflict state
was found.

\begin{figure}[t]
\begin{center}
\epsfig{file=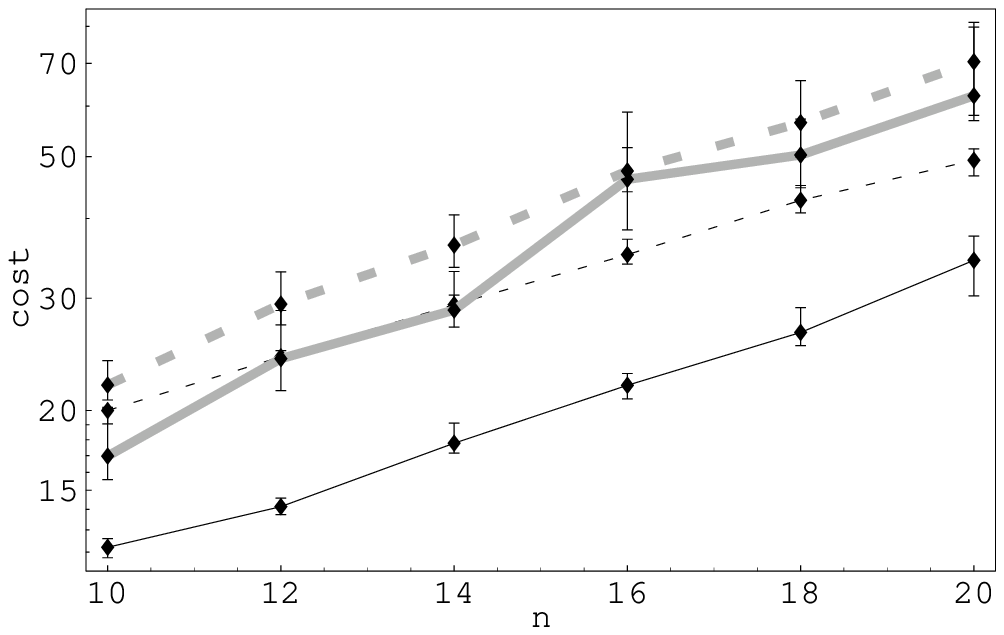,width=4in}
\end{center}
\caption{\label{SAT scaling}Median number of steps to find a state with the minimum number of conflicts, $\Cost=j/\Pmin$, for random 3-SAT problems with $\nSAT$ variables and no solutions using $j=\nSAT$ steps for each trial. Solid curves show behavior for the quantum heuristic for $\mu=4$ (black) and 6 (gray). For comparison, the dashed curves show the corresponding behavior for the classical GSAT heuristic on the same problems.
The algorithm used $T_0=0.539298$, $T_1=3.5105$, $R_0=4$ and $R_1= -3.4$ for $\mu=4$ and $T_0=0.87$, $T_1=2.7$, $R_0=2.57$ and $R_1= -1.73$ for $\mu=6$. The error bars show the 95\% confidence interval for the median~\cite[p.~124]{snedecor67}. The values are based on 100 problems for each $\nSAT$.}
\end{figure}

A comparison of the search costs, as measured by the expected number
of steps, for GSAT and the quantum heuristic is shown in \fig{SAT
scaling}.  Interestingly, the costs of the quantum algorithm are
comparable or below those of GSAT.  Actual search times for these
methods will depend on detailed implementations of the steps. Although
the number of elementary computational steps, involving evaluating the
number of conflicts in an assignment (and, in the case of GSAT, its
neighbors) are similar for both techniques, differences in the extent
to which operations can be optimized and the relative clock rates of
classical and quantum machines remain to be seen.

Since the trials of GSAT and this quantum heuristic are independent, 
decision problems allow a further quadratic speedup for
either of these techniques by combining them with amplitude
amplification~\cite{brassard98}.  Thus an interesting direction for
future work is the extent to which such techniques, extended to use
the number of conflicts in the search states, could give further
improvement for optimization problems as well.

%----------------------------------------------------------
\section{Traveling Salesman Problem}

The asymmetric traveling salesman problem (ATSP) has $\nTSP$ cities
with the distance from city $x$ to $y$ not necessarily equal to the
distance from $y$ to $x$, as would be approriate, for instance, in
planning routes along many one-way roads.  The goal is a
minimum-distance tour that visits every city exactly once and returns
to the starting point.  Many real-world planning and scheduling
problems can be modeled as TSP's~\cite{lawler85,miller91}.  An
$\nTSP$-city problem has $(\nTSP-1)!$ distinct tours starting and
ending in a given city, and the time required to solve ATSP grows
exponentially with $\nTSP$~\cite{garey79}.

We consider the average behavior for a class of TSPs studied in the
context of phase transitions in search
behavior~\cite{cheeseman91,zhang95,gent97}. Specifically, we looked at
problems with intercity distances picked independently from a normal
distribution with average $\mu$ and standard deviation $\sigma$, and
then rounded to the nearest integer.  Thus the probability a
particular tour has length $\tourlength$ is a discretized normal distribution
with average $\nTSP \mu$ and standard deviation $\sqrt{\nTSP}\,
\sigma$. The value of $\mu$ just sets the distance scale: we take it
equal to 100.

\subsection{Algorithm}

Quantum computers with $\nBits$ bits are most naturally used with
superpositions of $2^\nBits$ classical states.  Unlike the SAT
problem, ATSP has no direct mapping between the $(\nTSP-1)!$ search
states and the $2^\nBits$ states representable by superpositions of
$\nBits$ quantum bits.  Thus an important aspect of designing a
quantum algorithm for this problem is selecting a suitable
representation for the tours with binary elements.

One possibility, used in neural network approaches to
TSP~\cite{hopfield85}, represents each tour by a permutation matrix,
i.e., an $\nTSP \times \nTSP$ matrix of binary values. Specifically,
entry $i,j$ is one when city $i$ is at position $j$ in the tour. Thus
a tour gives exactly one entry in each row and column that is equal to
one, and the rest are zero. Since the choice of the first city is
arbitrary, we can take city 1 to be the first in each tour, leaving a
reduced $(\nTSP-1) \times (\nTSP-1)$ matrix describing the permutation
of cities 2 through $\nTSP$. Such a matrix can be represented with
$\nBits = (\nTSP-1)^2$ binary values. While this representation has a
fairly simple correspondence with the tours, considering
superpositions of all $2^\nBits$ possible values introduces many
states that do not correspond to tours (i.e., cases in which the
corresponding matrix has two or more 1's in a single row or column and
thus is not a permutation). Moreover, from a practical viewpoint, the
quadratic growth in the number of bits with $\nTSP$ severely limits
the problem sizes feasible for classical simulation. Thus, while this
representation may be useful for theoretical analyses and may provide
good performance, it is of limited use for empirically evaluating
quantum heuristics on classical machines.

Another representation, requiring fewer bits, simply enumerates the
tours starting from a given city in lexicographical order and
associates each with a bit string representing its index in this list.
For example, a 4-city problem with cities $A$, $B$, $C$, and $D$ has 6
distinct tours that start and end at city $A$: $ABCDA$, $ABDCA$,
$ACBDA$, $ACDBA$, $ADBCA$, and $ADCBA$.  Three bits are needed to
represent these 6 tours, ranging from 000 to 101.  The three bits give
8 possible values, so we also get 2 states without corresponding
tours: 110 and 111. 

Importantly for its use in a quantum algorith,  the permutation 
corresponding to a given index can be computed efficiently as a
particular example of techniques for ranking a variety of
combinatorial structures~\cite{nijenhuis78}. Specifically, consider 
index $i$, ranging from 0 to $(\nTSP-1)!-1$, as specifying a permutation 
of cities 2 through $\nTSP$ (with the understanding that the overall tour 
starts and ends with city 1). The value $\floor{i/(\nTSP-2)!}+2$, 
ranging from 2 to $\nTSP$, gives the first city in the permutation, 
and $i \bmod (\nTSP-2)!$ is the index of the permutation for the remaining
cities. Repeating this procedure once for each city gives the full 
permutation in $O(\nTSP)$ operations.

This representation uses a number of states that is the
closest power of two larger than or equal to the number of tours,
i.e., $2^\nBits \ge (\nTSP-1)!$, so $\nBits = \ceiling {\log_2
(\nTSP-1)!} \sim \nTSP \log_2 (\nTSP/e)$. This introduces some extra
states, not corresponding to tours, but far fewer than using the
permutation matrix representation. The algorithm must operate so as to
avoid giving much amplitude to these extra states.

For this problem, the phase adjustments depend on $c$, the scaled tour
length: $c(r) = \tourlength(r)/(\nTSP \mu)$ (where $\nTSP \mu$ is the
average tour length in this class of problems and $\tourlength(r)$ is
the length of tour $r$, i.e., the sum of costs between successive
cities in its path).  Smaller values of $c$ correspond to shorter
tours.  This definition makes the behavior of this algorithm similar
to that of the SAT algorithms described earlier, where $c$ represented
the number of conflicts in an assignment.  Since we want the extra
states (the ones added to make the total number of states a power of
two) to have small amplitudes, they should be assigned a large value
of $c$.  We chose $c=2$ for these extra states, so they are treated as
additional tours with especially long lengths.

With this representation of the tours, the Hamming distance between
two bit-strings has no simple relation to the difference of the
corresponding tours. This precludes a simple expression for the
conditional probability $P(c|d,c')$ used in the approximate
theoretical approach to identifying good phase parameters.  Instead,
we examined the performance for a random sample of problems, allowing
for different values of these parameters at each step of the
algorithm. We found only a small effect on performance from allowing
$\tau_h$ to vary from one step to the next, so we take it to be a
fixed value $\tau$ for the results reported below. However, using a
different $\rho_h$ value for each step significantly improved
performance. In particular, we took the values to vary linearly, as
with the SAT algorithm, i.e., we took $\rho_h = \rhoInit + \rhoRate h$
for $h$ from 1 to $j$ where $\rhoInit$ and $\rhoRate$ were new
parameters depending only on the class of problems, defined by $\mu$
and $\sigma$, and total number of steps.

In summary, for given choices of $\mu$, $\sigma$ and total number of
steps $j$, we selected parameters $\tau$, $\rhoInit$ and $\rhoRate$
giving the best average performance based on evaluation with a sample
of problems. The best choices we found for these parameters are listed
and discussed in the next section.

\subsection{Behavior}

\begin{figure}[t]
\begin{center}
\epsfig{file=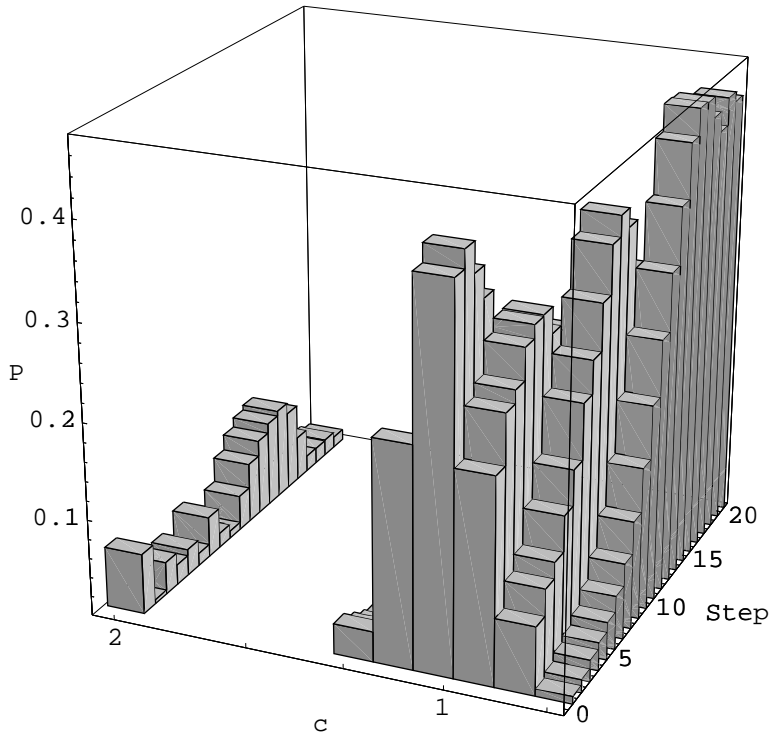,width=4in}
\end{center}
\caption{\label{sample prob}
Probability to find a tour vs.~scaled length $c$ and number of steps
for a 6-city ATSP with $\mu=100$ and $\sigma=40$. Tours are binned
according to their scaled lengths.  The minimum tour, with $c=0.72$,
is the only one in the bin with the smallest $c$ value. The bin at
$c=2$ has the extra states that do not correspond to tours.}
\end{figure}

\fig{sample prob} illustrates the algorithm's behavior for a random 
6-city problem.  The height of each bin is the sum of the
probabilities $p(s)$ of all the tours $s$ whose lengths fall in the
bin's range.  The initial step shows the initial distribution of tour
lengths.  We see the algorithm shifts amplitude from the longer tours
towards shorter ones.  For the problem instance shown in the figure,
the probability of finding the optimal tour peaks at 0.45 after 16
steps. After 20 steps, the bins with large probabilities correspond to
short tours. Thus when the algorithm does not find the optimal tour,
it will most likely still produce a tour close to optimal.  A similar
shift toward shorter tours occurs for other problems including those
with more cities.

To evaluate the average behavior for random 6- and 7-city problems we
generated 100 problems with $\mu=100$ and $\sigma$ equal to 5, 10, 15,
20, 30, and 40, and then applied the algorithm for 20 steps. The
results show the cost of solving the problems doesn't depend on the
$\sigma$ parameter of the distribution.  In other words, all problems
of a given size are equally complex; there are no hard and easy cases
for this algorithm.  This behavior differs from most classical
heuristics that work particularly well on under- and overconstrained
problems~\cite{gent97}.  This difference indicates our algorithm is
not taking full advantage of the problem structure, most likely due to
the simple form of the mixing matrix and the representation of
tours. It suggests an opportunity for improved performance by
developing mixing matrices using problem structure, or using state
representations where Hamming distance is a meaningful measure of tour
similarity.

Quantitatively, after 20 steps a solution to 6-city problems is found
with probability of roughly 30\%. This decreases to about 11\% for
7-city problems.  Thus, while the size of the search space increases
seven-fold, the cost goes up just by a factor of 3.  With additional
steps, the solution probability increases.  For example, after 30
steps (with different optimal choices for the parameters $\tau$,
$\rhoInit$ and $\rhoRate$), a solution to an average 7-city problem
with $\sigma = 40\%$ is found with a probability of about
16\%. Identifying the best number of steps, on average, remains an
open question. In particular, based on the behavior of the algorithm
for SAT, taking the number of steps proportional to $\nTSP$ may give
better performance with suitable choices for the phase parameters.

\begin{table}
\begin{center}
\begin{tabular}{|l|cccccc|}	\hline
deviation	& 5\%	& 10\%	& 15\% 	& 20\% 	& 30\%	& 40\% \\ \hline\hline

\multicolumn{7}{c}{\bf 6 cities} \\ \hline
$\rhoInit$	& .32	& .28	& .36	& .32	& .32	& .32  \\ \hline
$\rhoRate$	& .84	& .44	& .32	& .24	& .16	& .12  \\ \hline
$\tau$	 	& .12	& .12	& .12	& .12	& .12	& .12  \\ \hline

\multicolumn{7}{c}{\bf 7 cities} \\ \hline
$\rhoInit$	& .36	& .32	& .36	& .36	& .36	& .32  \\ \hline
$\rhoRate$	& .84	& .44	& .36	& .24	& .16	& .12  \\ \hline
$\tau$		& .12	& .12	& .12	& .12	& .12	& .12  \\ \hline

\end{tabular}
\end{center}
\caption{\label{cities}Optimal parameters for 20 steps.}
\end{table}

\begin{figure}[t]
\begin{center}
\epsfig{file=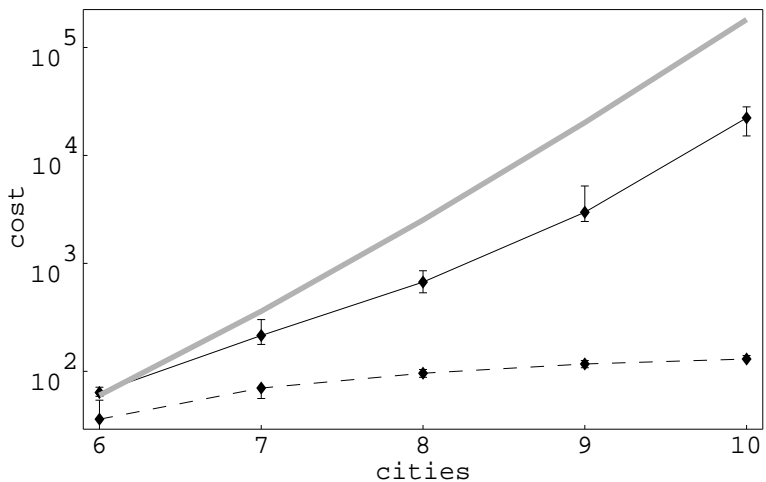,width=4in}
\end{center}
\caption{\label{scaling}
Median number of steps to find the optimal tour vs.~$\nTSP$
for $\sigma=40\%$. The black curve shows the quantum search cost, $20/\Pmin$, using the optimal parameters for 6 and 7 cities,
which are identical in this case. The gray curve shows the scaling of random selection and the dashed curve is the cost estimate for the classical method DFBnB described in the text. The error bars show the 95\% confidence intervals for the medians. Each point is based on a sample of 100
problems.}
\end{figure}

\tbl{cities} summarizes the the phase parameters used in 6 and 7-city 
simulations, i.e., values of $\rhoInit$, $\rhoRate$ and $\tau$ that
maximize the probability of finding the optimal tour in 20 steps. The
optimal $\rho^{(h)} = \rhoInit + \rhoRate h$ values are almost
independent of the problem size, but are largely (and very
consistently) dependent on $\sigma$: doubling $\sigma$ roughly halves
the value of $\rho$.  This is to be expected because doubling $\sigma$
also doubles the range of scaled tour lengths, $c-1$ (where 1 is the
average value of $c$ for the class of problems). Thus halving
$\rho$-values keeps the phase matrix entries $p_{c(r)}$ about the
same, up to an irrelevant overall phase.  Since the mixing matrix does
not use the problem structure, it is not surprising that the optimal
value of the mixing matrix constant $\tau$ is independent of $\nTSP$
and $\sigma$ for a given number of steps.

\fig{scaling} shows the scaling behavior for a fixed choice of $\sigma$. 
For $\nTSP>7$, the simulations are too slow to allow finding optimal
parameter values. Instead, noting the parameters for the 6 and 7 city
problems, given in \tbl{cities}, were the same for $\sigma=40\%$, we
continued using these parameters for the larger $\nTSP$.  The solution
probability after 20 steps decreases as $e^{-1.2 \nTSP}$. These larger
cases continue to show the shift of probability toward short tours.
The resulting costs grow more slowly than the expected number of
trials for random selection to find the optimal tour, $(\nTSP-1)!/2$.
This random selection contrasts with exhaustive enumeration, with cost
$(\nTSP-1)!$, which not only finds the optimum but also guarantees the
result is indeed optimal.

As another comparison, \fig{scaling} also shows an estimate of the
median cost for a good classical method, depth-first branch-and-bound
(DFBnB)~\cite{zhang00}.  This technique relies on the assignment
problem (AP), in which each city is linked to another so that the
total cost of these links is minimized. The resulting links need not
form a complete tour, in which case the search proceeds by considering
subproblems in which some of the links in the AP solution are not
allowed. The initial AP can be solved in time of order $\nTSP^3$ while
subsequent instances appearing in the subproblems require only
$\nTSP^2$ operations. Ignoring overall constants and the, usually
relatively minor, cost to find an initial upper bound on tour lengths
by adjusting the result of the initial AP~\cite{karp79}, we take the
search cost associated with this method to be $C_{\rm classical}
\approx \nTSP^3 + b \nTSP^2$ where $b$ is the number of subproblems
evaluated during the search. To compare with the quantum search cost
measured in terms of steps, note that each step involves computing the
cost of tours in superposition, which requires $O(\nTSP)$ operations. Thus
we divide $C_{\rm classical}$ by $\nTSP$ to obtain the cost estimates
used for comparison in \fig{scaling}. While the estimate does not include
multiplicative constants, it does show the classical heuristic grows
much more slowly than the quantum method introduced here. One caveat
is that for large problems $b$ grows exponentially but for the small
problems accessible to the quantum simulations, i.e., $\nTSP$ up to
about 10, about a third of the instances are solved without any search
since the initial assignment problem returns a complete tour. Many
other cases are solved by expanding just a few subproblems. Thus many
of these small problems are dominated by the cost of the initial AP
and the figure does not show the eventual exponential growth in
cost. We should also note this classical algorithm, unlike the quantum
algorithm and GSAT, is complete and guarantees its result is indeed
the minimum.

In spite of these limitations, this comparison does suggest the
quantum technique is not particularly effective in exploiting problem
structure, especially when compared with the satisfiability search
described above.  An interesting open question is whether this
indicates quantum heuristics are inherently less effective for ATSP
than for SAT. Other possible reasons for the relatively large costs
with the quantum algorithm include the choice of problem
representation, which does not explicitly use the relations among
tours with many common edges, and nonoptimal values of phase
adjustment parameters and number of steps for the larger problem sizes
considered here.

%----------------------------------------------------------
\section{Conclusion}

A quantum algorithm shifting amplitude toward low-cost states is
effective for combinatorial optimization problems, and does not
require prior knowledge of the minimum cost for particular instances.
This work illustrates how quantum techniques developed for decision
problems can also apply to optimization, but only if they make use of
the cost structure of the states.  The experiments for satisfiability
indicate appropriate phase parameters can allow the algorithm to have
performance at least comparable to classical heuristics.  The
exponential increase in cost for classical simulation precludes
evaluating these observations with larger problems.

For the asymmetric traveling salesman problem, our algorithm works
fairly well for small instances, and specifically is much better than
random selection.  Its performance is independent of the standard
deviation of the intercity distances.  By allowing contributions from
different search states to interfere, the algorithm avoids the large
resource costs of using quantum parallelism without mixing the
amplitudes~\cite{cerny93}.  One direction for future work is examining
other ways of encoding the problem.  Currently all the tours are
enumerated in lexicographical order, and the mixing matrix doesn't
take advantage of the problem structure.  Other enumerations may allow
the mixing matrix to incorporate some problem structure.  In analogy
with the mixing of assignments in SAT, we could favor mixing
amplitudes amongst similar tours, since tours that have a large number
of edges in common are likely to have similar lengths. Ideally, such a
representation could provide simple analytic evaluation of the
conditional probability and thereby suggest better phase parameters
for larger problems.

For some classical search methods applied to decision problems, the
method can be incorporated in a quantum algorithm to give a further
improvement~\cite{brassard98,cerf98}. Hence an interesting open
question is whether such techniques can generalize to optimization
problems and thereby improve, for instance, on GSAT and the
branch-and-bound examples of classical heuristics discussed here.

Unlike decision problems where results are easily verified, this
optimization algorithm's results cannot be directly checked for
optimality. Thus, as with classical heuristics, such as GSAT and
simulated annealing~\cite{kirkpatrick83}, applied to optimization
problems, this algorithm does not indicate whether its result is
indeed optimal. Moreover, although we focus on the probability to find
the optimal tour, algorithms for optimization problems are
characterized more generally by their trade-off between search cost
and quality of the result. Such trade-offs may be particularly
relevant for implementations of quantum computers limited to
relatively few steps due to decoherence. For the algorithms considered
here, the required number of coherent computational steps (i.e., the
length of a single trial) grows at most linearly with with problem
size. This contrasts with the exponentially growing number of steps in
a single trial of amplitude amplification. Thus the structure-based
algorithms make less stringent requirements on the extent to which
coherence can be maintained.

In summary, we have shown how to use the cost associated with states
in an optimization problem to adjust a superposition to increase the
amplitudes associated with low-cost states. This opens a new direction
for applying quantum computers to combinatorial searches, but the
extent to which this capability can improve on classical heuristics,
on average, remains an open question.

%----------------------------------------------------------
\small
\section*{\small Acknowledgments}

We have benefited from discussions with Wolf Polak, Eleanor Rieffel
and Christof Zalka. We also thank Weixiong Zhang for providing his
TSP search program and helpful comments on its behavior.

%\bibliography{refs,physrev}\bibliographystyle{plain}

\begin{thebibliography}{10}

\bibitem{boyer96}
Michel Boyer, Gilles Brassard, Peter Hoyer, and Alain Tapp.
\newblock Tight bounds on quantum searching.
\newblock In T.~Toffoli et~al., editors, {\em Proc. of the Workshop on Physics
  and Computation (PhysComp96)}, pages 36--43, Cambridge, MA, 1996. New England
  Complex Systems Institute.

\bibitem{brassard98}
Gilles Brassard, Peter Hoyer, and Alain Tapp.
\newblock Quantum counting.
\newblock In K.~Larsen, editor, {\em Proc. of 25th Intl. Colloquium on
  Automata, Languages, and Programming (ICALP98)}, pages 820--831, Berlin,
  1998. Springer.
\newblock {Los Alamos} preprint quant-ph/9805082.

\bibitem{cerf98}
Nicolas~J. Cerf, Lov~K. Grover, and Colin~P. Williams.
\newblock Nested quantum search and {NP}-complete problems.
\newblock In {\em Applicable Algebra in Engineering, Communication and
  Computing}. Springer, Berlin, 1998.
\newblock {Los Alamos} preprint quant-ph/9806078.

\bibitem{cerny93}
Vladimir Cerny.
\newblock Quantum computers and intractable ({NP}-complete) computing problems.
\newblock {\em Physical Review A}, 48:116--119, 1993.

\bibitem{cheeseman91}
Peter Cheeseman, Bob Kanefsky, and William~M. Taylor.
\newblock Where the really hard problems are.
\newblock In J.~Mylopoulos and R.~Reiter, editors, {\em Proceedings of
  IJCAI91}, pages 331--337, San Mateo, CA, 1991. Morgan Kaufmann.

\bibitem{deutsch85}
D.~Deutsch.
\newblock Quantum theory, the {Church-Turing} principle and the universal
  quantum computer.
\newblock {\em Proc. R. Soc. London A}, 400:97--117, 1985.

\bibitem{divincenzo95}
David~P. DiVincenzo.
\newblock Quantum computation.
\newblock {\em Science}, 270:255--261, 1995.

\bibitem{freuder92}
Eugene~C. Freuder and Richard~J. Wallace.
\newblock Partial constraint satisfaction.
\newblock {\em Artificial Intelligence}, 58:21--70, 1992.

\bibitem{garey79}
M.~R. Garey and D.~S. Johnson.
\newblock {\em Computers and Intractability: A Guide to the Theory of
  NP-Completeness}.
\newblock W. H. Freeman, San Francisco, 1979.

\bibitem{gent97}
Ian~P. Gent, Ewan MacIntyre, Patrick Prosser, and Toby Walsh.
\newblock The scaling of search cost.
\newblock In {\em Proc. of the 14th Natl. Conf. on Artificial Intelligence
  (AAAI97)}, pages 315--320, Menlo Park, CA, 1997. AAAI Press.

\bibitem{grover96}
Lov~K. Grover.
\newblock Quantum mechanics helps in searching for a needle in a haystack.
\newblock {\em Physical Review Letters}, 78:325--328, 1997.
\newblock {Los Alamos} preprint quant-ph/9706033.

\bibitem{grover97b}
Lov~K. Grover.
\newblock Quantum search on structured problems.
\newblock {\em Chaos, Solitons, and Fractals}, 10:1695--1705, 1999.

\bibitem{hogg98}
Tad Hogg.
\newblock Highly structured searches with quantum computers.
\newblock {\em Physical Review Letters}, 80:2473--2476, 1998.
\newblock Preprint at
  pub\-lish.aps.org/eprint/gate\-way/eplist/aps1997oct30\_002.

\bibitem{hogg98e}
Tad Hogg.
\newblock Single-step quantum search using problem structure.
\newblock {Los Alamos} preprint quant-ph/9812049, 1998.

\bibitem{hogg00}
Tad Hogg.
\newblock Quantum search heuristics.
\newblock {\em Physical Review A}, 61:052311, 2000.
\newblock Preprint at
  pub\-lish.aps.org/eprint/gate\-way/eplist/aps1999oct19\_002.

\bibitem{hogg96d}
Tad Hogg, Bernardo~A. Huberman, and Colin~P. Williams, editors.
\newblock {\em Frontiers in Problem Solving: Phase Transitions and Complexity},
  volume~81, Amsterdam, 1996. Elsevier.
\newblock Special issue of {\it Artificial Intelligence}.

\bibitem{hogg98b}
Tad Hogg, Carlos Mochon, Eleanor Rieffel, and Wolfgang Polak.
\newblock Tools for quantum algorithms.
\newblock {\em Intl. J. of Modern Physics C}, 10:1347--1361, 1999.
\newblock {Los Alamos} preprint quant-ph/9811073.

\bibitem{hopfield85}
J.~J. Hopfield and D.~W. Tank.
\newblock ``{Neural}'' computation of decisions in optimization problems.
\newblock {\em Biol. Cybern}, 52:141--152, 1985.

\bibitem{karp79}
Richard~M. Karp.
\newblock A patching algorithm for the nonsymmetric traveling-salesman problem.
\newblock {\em SIAM J. on Computing}, 8:561--573, 1979.

\bibitem{kirkpatrick83}
S.~Kirkpatrick, C.~D. Gelatt, and M.~P. Vecchi.
\newblock Optimization by simulated annealing.
\newblock {\em Science}, 220:671--680, 1983.

\bibitem{kirkpatrick94}
Scott Kirkpatrick and Bart Selman.
\newblock Critical behavior in the satisfiability of random boolean
  expressions.
\newblock {\em Science}, 264:1297--1301, 1994.

\bibitem{lawler85}
E.~L. Lawler, J.~K. Lenstra, A.~H. G.~Rinnooy Kan, and D.~B. Shmoys, editors.
\newblock {\em The Traveling Salesman Problem}.
\newblock John Wiley, NY, 1985.

\bibitem{miller91}
Donald~L. Miller and Joseph~F. Pekny.
\newblock Exact solution of large asymmetric traveling salesman problems.
\newblock {\em Science}, 251(4995):754--761, Feb. 15 1991.

\bibitem{nijenhuis78}
A.~Nijenhuis and H.~S. Wilf.
\newblock {\em Combinatorial Algorithms for Computers and Calculators}.
\newblock Academic Press, New York, 2nd edition, 1978.

\bibitem{selman92}
Bart Selman, Hector Levesque, and David Mitchell.
\newblock A new method for solving hard satisfiability problems.
\newblock In {\em Proc. of the 10th Natl. Conf. on Artificial Intelligence
  (AAAI92)}, pages 440--446, Menlo Park, CA, 1992. AAAI Press.

\bibitem{snedecor67}
George~W. Snedecor and William~G. Cochran.
\newblock {\em Statistical Methods}.
\newblock Iowa State Univ. Press, Ames, Iowa, 6th edition, 1967.

\bibitem{spector99}
Lee Spector, Howard Barnum, Herbert~J. Bernstein, and Nikhil Swamy.
\newblock Finding a better-than-classical quantum {AND/OR} algorithm using
  genetic programming.
\newblock In P.~Angeline, editor, {\em Proc. of the 1999 Congress on
  Evolutionary Computing}, Washington, DC, 1999. IEEE.

\bibitem{zhang00}
Weixiong Zhang.
\newblock Depth-first branch-and-bound versus local search: A case study.
\newblock In {\em Proc. of the 17th Natl. Conf. on Artificial Intelligence
  (AAAI2000)}. AAAI, 2000.

\bibitem{zhang95}
Weixiong Zhang and Richard~E. Korf.
\newblock A study of complexity transitions on the asymmetric traveling
  salesman problem.
\newblock {\em Artificial Intelligence}, 81:223--239, 1996.

\end{thebibliography}

\end{document}